\documentclass[twocolumn,trackchanges]{aastex62}
\usepackage[section]{placeins}
\usepackage{longtable}
\usepackage{hyperref}
\usepackage{url}
\usepackage{amsmath}
\graphicspath{{./}{figures/}}
\newcommand{\ra}[4]{#1\textsuperscript{h}#2\textsuperscript{m}#3\fs#4}
\newcommand{\dec}[4]{$#1$\textsuperscript{d}#2\textsuperscript{m}#3\fs#4}

\usepackage[hang,flushmargin]{footmisc} 
\usepackage{url}

\newcommand{\Einstein}{\altaffiliation{Einstein Fellow}}
\newcommand{\Hubble}{\altaffiliation{Hubble Fellow}}
\newcommand{\CfA}{\affiliation{Center for Astrophysics \textbar{} Harvard \& Smithsonian, 60 Garden Street, Cambridge, MA 02138-1516, USA}}

\newcommand{\Carnegie}{\affiliation{Observatories of the Carnegie Institute for Science, 813 Santa Barbara Street, Pasadena, CA 91101-1232, USA}}

\newcommand{\Edinburgh}{\affiliation{Institute for Astronomy, University of Edinburgh, Royal Observatory, Blackford Hill EH9 3HJ, UK}}
\newcommand{\Birmingham}{\affiliation{Birmingham Institute for Gravitational Wave Astronomy and School of Physics and Astronomy, University of Birmingham, Birmingham B15 2TT, UK}}
\newcommand{\CIERA}{\affiliation{Center for Interdisciplinary Exploration and Research in Astrophysics and Department of Physics and Astronomy, \\Northwestern University, 2145 Sheridan Road, Evanston, IL 60208-3112, USA}}

\newcommand{\Ohio}{\affiliation{Astrophysical Institute, Department of Physics and Astronomy, 251B Clippinger Lab, Ohio University, Athens, OH 45701-2942, USA}}
\newcommand{\AAS}{\affiliation{American Astronomical Society, 1667 K~Street NW, Suite 800, Washington, DC 20006-1681, USA}}

\shorttitle{Magellan Follow-up of S190814bv}
\shortauthors{Gomez et al.}

\begin{document}

\title{A Galaxy-Targeted Search for the Optical Counterpart of the Candidate NS--BH Merger S190814bv with \textit{Magellan}}

\correspondingauthor{Sebastian Gomez}
\email{sgomez@cfa.harvard.edu}

\author[0000-0001-6395-6702]{S.~Gomez}
\CfA

\author[0000-0002-0832-2974]{G.~Hosseinzadeh}
\CfA

\author[0000-0002-2478-6939]{P.~S.~Cowperthwaite}
\Hubble\Carnegie

\author[0000-0002-5814-4061]{V.~A.~Villar}
\CfA

\author[0000-0002-9392-9681]{E.~Berger}
\CfA

\author[0000-0002-3003-3183]{T.~Gardner}
\affiliation{Astronomy Department, University of Michigan, Ann Arbor, MI 48109, USA}

\author[0000-0002-8297-2473]{K.~D.~Alexander}
\Einstein\CIERA

\author[0000-0003-0526-2248]{P.~K.~Blanchard}
\CIERA

\author[0000-0002-7706-5668]{R.~Chornock}
\Ohio

\author{M.~R.~Drout}
\Carnegie\affiliation{Department of Astronomy and Astrophysics, University of Toronto, 50 St.~George St., Toronto, Ontario, M5S 3H4 Canada}

\author[0000-0003-0307-9984]{T.~Eftekhari}
\CfA

\author[0000-0002-7374-935X]{W.~Fong}
\CIERA

\author[0000-0003-4341-9824]{K.~Gill}
\CfA

\author[0000-0003-4768-7586]{R.~Margutti}
\CIERA

\author[0000-0002-2555-3192]{M.~Nicholl}
\Edinburgh\Birmingham

\author[0000-0001-8340-3486]{K.~Paterson}
\CIERA

\author[0000-0003-3734-3587]{P.~K.~G.~Williams}
\CfA\AAS


\begin{abstract}
On 2019 August 14 the Laser Interferometer Gravitational Wave Observatory (LIGO) and the Virgo gravitational wave interferometer announced the detection of a binary merger, S190814bv, with a low false alarm rate (FAR) of about 1 in $1.6\times 10^{25}$ yr, a distance of $267\pm 52$ Mpc, a 90\% (50\%) localization region of about 23 (5) deg$^2$, and a probability of being a neutron star--black hole (NS--BH) merger of $>99\%$. The LIGO/Virgo Collaboration (LVC) defines NS--BH such that the lighter binary member has a mass of $<3$ M$_\odot$ and the more massive one has $>5$ M$_\odot$, and this classification is in principle consistent with a BH--BH merger depending on the actual upper mass cutoff for neutron stars. Additionally, the LVC designated a probability that the merger led to matter outside the final BH remnant of $<1\%$, suggesting that an electromagnetic (EM) counterpart is unlikely. Here we report our optical follow-up observations of S190814bv using the \textit{Magellan} Baade 6.5 m telescope to target all 96 galaxies in the Galaxy List for the Advanced Detector Era catalog within the 50\% localization volume (representing about 70\% of the integrated luminosity within this region).  No counterpart was identified to a median $3\sigma$ limiting magnitude of $i=22.2$ ($M_i\approx -14.9$ mag), comparable to the brightness of the optical counterpart of the binary neutron star merger GW170817 at the distance of S190814bv; similarly, we can rule out an on-axis jet typical of short GRBs.  However, we cannot rule out other realistic models, such as a kilonova with only $\sim 0.01$ M$_\odot$ of lanthanide-rich material, or an off-axis jet with a viewing angle of $\theta_{\rm obs}\gtrsim 15^\circ$.
\end{abstract}


\section{Introduction}
\label{sec:intro}

\begin{figure*}[!t]
\begin{center}
\scalebox{1.}
\centering
{\includegraphics[width=0.9\textwidth]{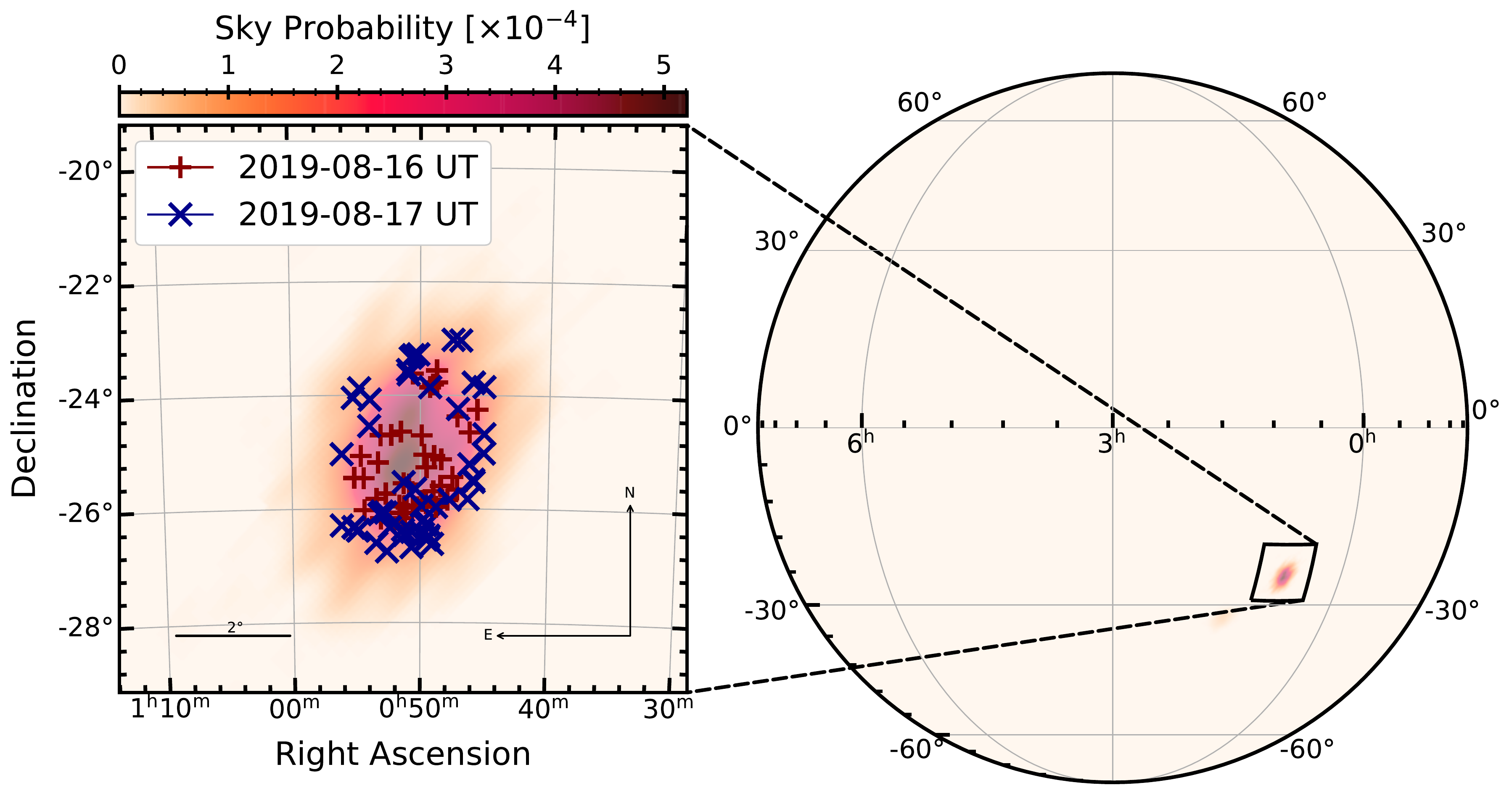}}
\caption{Localization region of S190814bv overlaid with the locations of galaxies observed in our Magellan search on 2019 August 16 (red) and 17 (blue).  These represent all galaxies present in the GLADE catalog within the 50\% localization volume, corresponding to about 50\% of all galaxies in this region down to a luminosity of 0.15 $L^*$, and about 75\% of the integrated galaxy luminosity within the region. 
\label{fig:skymap}}
\end{center}
\end{figure*}

Observing run 3 (O3) of Advanced LIGO and Virgo (ALV) commenced on 2019 April 1, with a 50\% increase in sensitivity compared to Observing Runs 1 and 2, corresponding to a binary neutron star merger detection range of about 125 Mpc \citep{LIGOLocalization}.  As of the end of 2019 July the LIGO/Virgo Collaboration (LVC) has issued 22 public alerts (that were not subsequently retracted), of which 18 are high probability binary black hole (BH--BH) mergers, one (S190425z) is a likely binary neutron star (NS--NS) merger, two are ambiguous in terms of their astrophysical nature and overall detection robustness (S190426c and S190510g), and one is likely terrestrial (non-astrophysical) in origin (S190718y).  While the candidate event S190426c was initially considered the first possible NS--BH merger, its probability of being such an event is only 13\% and its false alarm rate (FAR) is only 1 in 1.6 yr (with a corresponding 14\% probability that it is terrestrial in origin).

On 2019 August 14 at 21:10:39 UTC (=58709.88240 MJD), ALV detected a GW candidate event, designated S190814bv, with an incredibly low FAR of 1 in $1.56\times10^{25}$ yr, a luminosity distance of $267\pm 52$ Mpc, and a 90\% (50\%) confidence localization region of 23 (5) deg$^2$ \citep{gcn25324,gcn25333}; see Figure~\ref{fig:skymap} for the localization map.  S190814bv was initially classified with 100\% confidence as a ``mass-gap'' event, namely a merger in which at least one of the binary members is in the mass range \mbox{$3-5$ M$_\odot$}.  About 13 hr later, initial parameter estimation revised the claimed nature of the event to NS--BH merger (i.e., a merger in which the lightest member has a mass of $<3$ M$_\odot$) with a probability of $>99\%$.  We note that this definition allows the event to actually be a BH--BH merger, depending on the actual upper mass cutoff for neutron stars.  In addition, the initial parameter estimation indicates that the probability for matter outside of the final merger remnant (the BH) is $<1\%$, suggesting that the merger is unlikely to produce electromagnetic (EM) emission \citep{gcn25333}.  We note that this probability encapsulates information from the parameters of the binary (e.g., mass ratio, spin), which are not currently publicly available, and assumptions about the neutron star equation of state.  

Here we report our optical follow-up of S190814bv using the \textit{Magellan} Baade 6.5 m telescope to target galaxies within the localization volume.  In \S\ref{sec:Magellan} we present our \textit{Magellan} observations. In \S\ref{sec:comp} we compare the results of our search to the kilonova emission of GW170817, to theoretical kilonova models, and to on- and off-axis afterglow models. We summarize and draw initial conclusions in \S\ref{sec:conc}.

\section{Galaxy--Targeted Follow-up with \textit{Magellan}}
\label{sec:Magellan}

\begin{figure*}[t!]
\begin{center}
\scalebox{1.}
\centering
{\includegraphics[width=\textwidth]{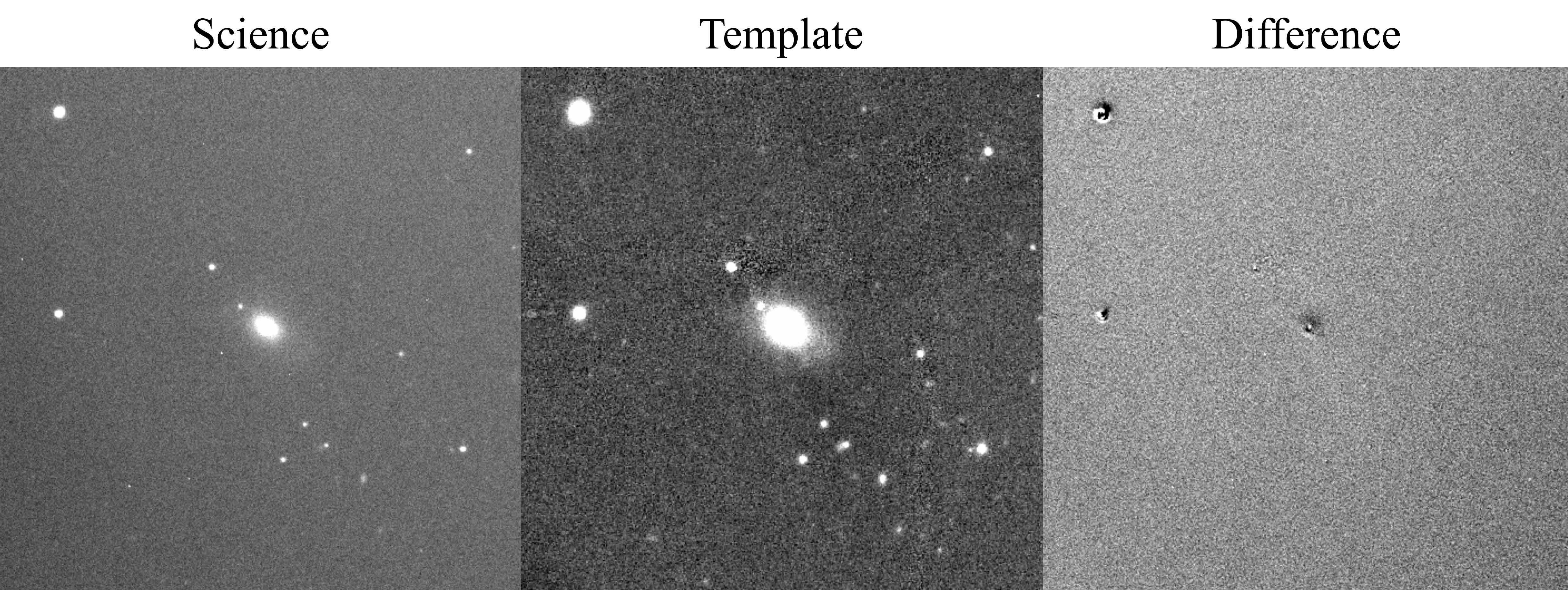}}
\vspace{0.0in}
\caption{Galaxy imaged with our \textit{Magellan} search in the field of S190814bv (left), along with the corresponding reference image from PS1 $3\pi$ (middle); \citealt{Chambers_pan-starrs1_2016}), and the resulting subtraction (right). The images are $2.5'$ on a side, oriented with north up and east to the left.  The difference image exhibits only astrometric noise and cosmic--ray artifacts; no counterpart is identified in this image to a $3\sigma$ limit of $i=22.2$ mag.
\label{fig:galaxy}}
\end{center}
\end{figure*}

Upon receipt of the LVC public alert for S190814bv, our automated software generated a list of galaxies from the Galaxy List for the Advanced Detector Era (GLADE) catalog \citep{dalya_glade:_2018} covering the 90\% confidence volume and ranked by probability within the volume (see \citealt{2019ApJ...880L...4H} for a detailed discussion).

We commenced follow-up observations with the Inamori--\textit{Magellan} Areal Camera and Spectrograph (IMACS) on the \textit{Magellan} Baade 6.5 m telescope at Las Campanas Observatory in Chile on 2019 August 16 at 08:15:38 UT (35 hr post-merger) and continued until morning twilight, with our last exposure ending at 10:30:02 UT, observing a total of 45 galaxies \citep{gcn25366}. On the following night (2019 August 17) we observed from 08:08:39 to 10:27:14 UT ($59-61.3$ hr post-merger) and imaged 51 additional galaxies \citep{gcn25382}; see Figure~\ref{fig:skymap} for the galaxy positions relative to the localization map of S190814bv.  We obtained a single 60-s $i$-band image per galaxy to minimize moonlight contamination.

The images were processed in real time following each exposure using a dedicated Python pipeline to perform bias subtraction and flatfielding. Image subtraction was performed relative to Pan-STARRS1 $3\pi$ $i$-band images using the HOTPANTS software \citep{Becker2015} and we searched for candidate transients through visual inspection. No transient sources were uncovered in these observations to a median $3\sigma$ limiting magnitude and 90\% percentile range of $i=22.2^{+0.3}_{-0.6}$. These limits were calculated for each individual image, from a measure of the average sky background and systematic noise sources, calibrated relative to field stars from the PS1 $3\pi$ catalog. An example of the \textit{Magellan} images and image subtraction results is shown in Figure~\ref{fig:galaxy}.  We provide the information for all of the individual galaxies in Table~\ref{tab:magellan} where the reported magnitudes have been corrected for negligible Galactic extinction with \mbox{$E(B-V)\approx 0.03$} \citep{Schlafly11}.

The 96 observed galaxies comprise all galaxies from the GLADE catalog in the 50\% confidence volume of S190814bv (\mbox{$1.8\times 10^4$ Mpc$^3$}) with luminosities of \mbox{$\gtrsim 0.15$ $L^*$}. To estimate the completeness of this sample we integrate the B-band galaxy luminosity function down to the same limit and estimate an expected 195 galaxies within the 50\% localization volume of \mbox{$1.8\times 10^4$ Mpc$^3$}. We adopt values of $M_* = -20.75$, $\phi_* = 0.0055$, and $\alpha = -1.20$ for the B band luminosity function $\phi\left(M\right) = 0.4 \ln{10} \phi^{*}10^{L\left(\alpha+1\right) }\exp{\left(-10^{L}\right)}$, where $L = 0.4\left(M^*-M\right)$ \citep{Faber07}. This indicates that in terms of number of galaxies our search was about 50\% complete within the 50\% confidence region, or equivalently that we covered about 25\% of the overall probability of the location of S190814bv. More importantly, in terms of integrated luminosity (and hence roughly stellar mass) the resulting overall fraction is higher, about 35\% of the probability (which comprises more than half of the integrated luminosity). We covered every galaxy in the GLADE catalog with \mbox{$L \geq 0.75 L^*$}, a total of 44 galaxies in the 50\% localization volume. From integrating B band luminosity function down to this luminosity we predict a total of 47 galaxies. Therefore, the GLADE catalog is essentially complete ($\approx95$\% coverage) within the 50\% volume down to a luminosity of 0.75 \textit{L}$^*$.

\section{Comparison to GW170817 and Theoretical Models}
\label{sec:comp}

\begin{figure*}[t!]
\begin{center}
\scalebox{1.}
\centering
{\includegraphics[width=\textwidth]{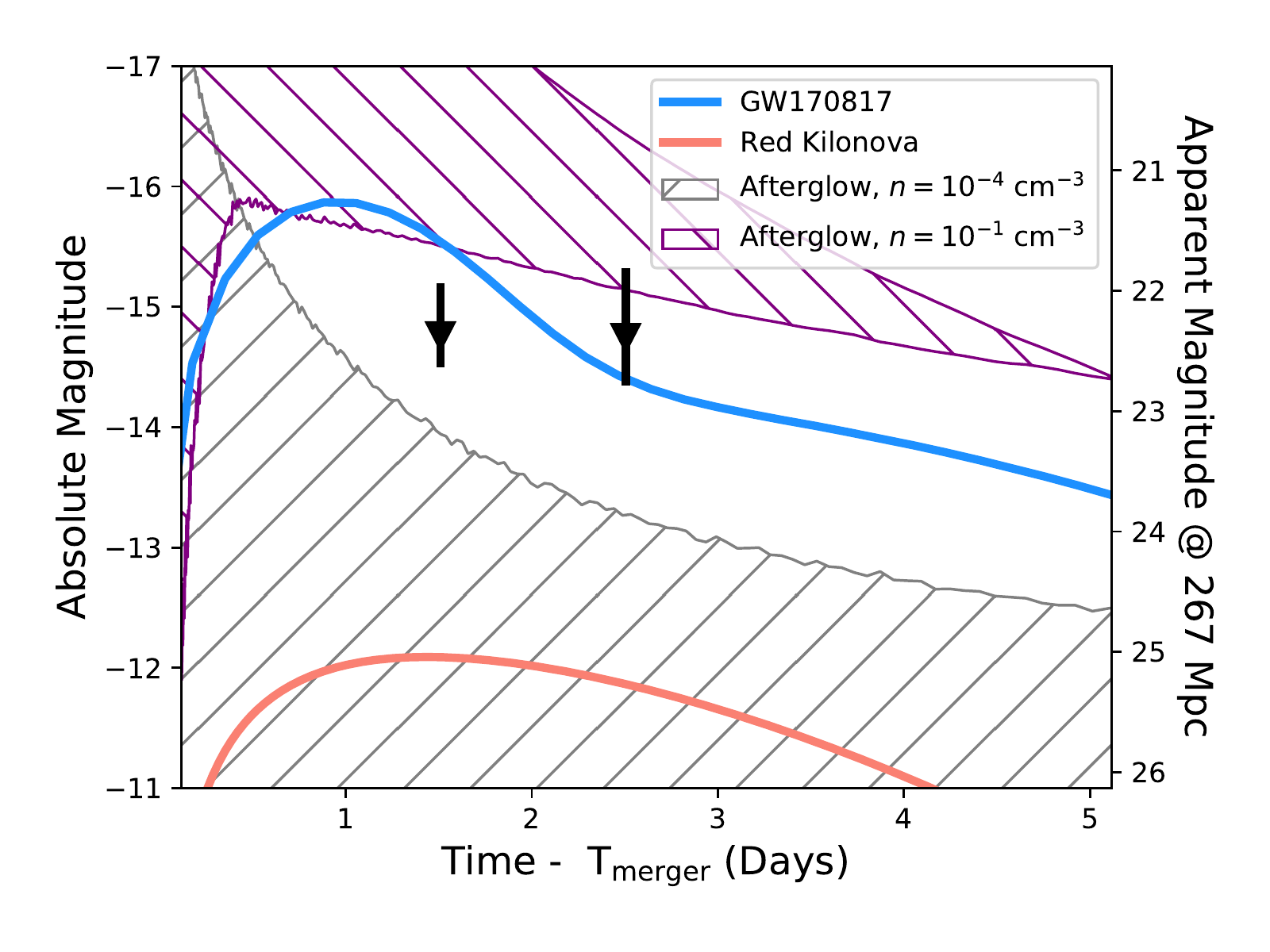}}
\vspace{-0.4in}
\caption{Our \textit{Magellan} search limits compared to several potential optical emission light curve models. The black arrows and lines represent the 90\% range for the limiting absolute magnitudes in each night, using the relevant distance of each targeted galaxy. Also shown are the model light curves for the kilonova associated with GW170817 (blue; \citealt{Villar+17b}), a lanthanide-rich kilonova with an ejecta mass of 0.01 $M_\odot$ (red), and afterglows based on a range of short GRB properties \citep{Fong+15} for high density ($n = 10^{-1}$ cm$^{-3}$; purple) and low density ($n = 10^{-4}$ cm$^{-3}$; gray) environments. The shaded regions represent the span of possible models for different viewing angles, from on-axis jets at the brightest edge of the shaded regions, to an off-axis jet ($15^\circ$ angle) at the bottom edge of the regions. The right-hand-side ordinate shows the relevant apparent magnitudes for the mean distance of S190814bv ($267$ Mpc) to provide an indication of the rough depth required to detect the various models.
\label{fig:lightcurve}}
\end{center}
\end{figure*}

At the 90\% confidence distance range of S190814bv a kilonova identical to that associated with GW170817 (which peaked at a magnitude of $M_i\approx -15.8$ mag; \citealt{Villar+17b}) would peak at a magnitude of $i\approx 20.5-22$ at a timescale of about 1 day post-merger; see Figure~\ref{fig:lightcurve}. In said figure we show the best-fit model of GW170817 from \citep{Villar+17b}. This is a three component kilonova model, with a ``blue" lanthanide-poor component, an intermediate ``purple" component, and a ``red" lanthanide-rich component; with a respective increasing opacity, spanning $\kappa = 0.5 - 10$ cm$^2$ g$^{-1}$, velocities spanning $v_{\rm ej} = 0.1$--$0.26$c, and a total mass of $M_{\rm ej} \approx 0.078$ M$_\odot$. The limiting magnitudes for our first night of observations span a median and 90\% confidence range that can rule out a GW170817-like kilonova, while the observations obtained on the following night rule out such a kilonova only for the lower half of the distance range.

The kilonova associated with GW170817 was dominated at early time by a bright blue emission component, possibly due to ejecta from the collision interface of the two neutron stars \citep{2014MNRAS.441.3444M,Nicholl+17a} or a (short-lived) hypermassive neutron star or magnetar remnant \citep{2018ApJ...869L...3F,2018ApJ...856..101M}.  In the case of an NS--BH merger such processes are not expected, and the emission will instead be dominated by dynamical ejecta or an accretion disk wind.  We therefore compare our limits to a model representative of only the ``red" lanthanide-rich component of GW170817 with a nominal ejecta mass of 0.01 M$_\odot$ and a high opacity of $\kappa = 10$ cm$^2$ g$^{-1}$, which has a peak brightness of $M_i\approx -13$ mag ($i\approx 23.5-25$ mag at the distance range of S190814bv) on a timescale of about $1-2$ days (Figure~\ref{fig:lightcurve}). We find that the resulting models are not significantly affected by the ejecta mass. Such a model is beyond the reach of our observations for any ejecta mass $< 0.03$ M$_\odot$.

\begin{figure}[h!]
\begin{center}
\scalebox{1.}
\centering
{\includegraphics[width=\columnwidth]{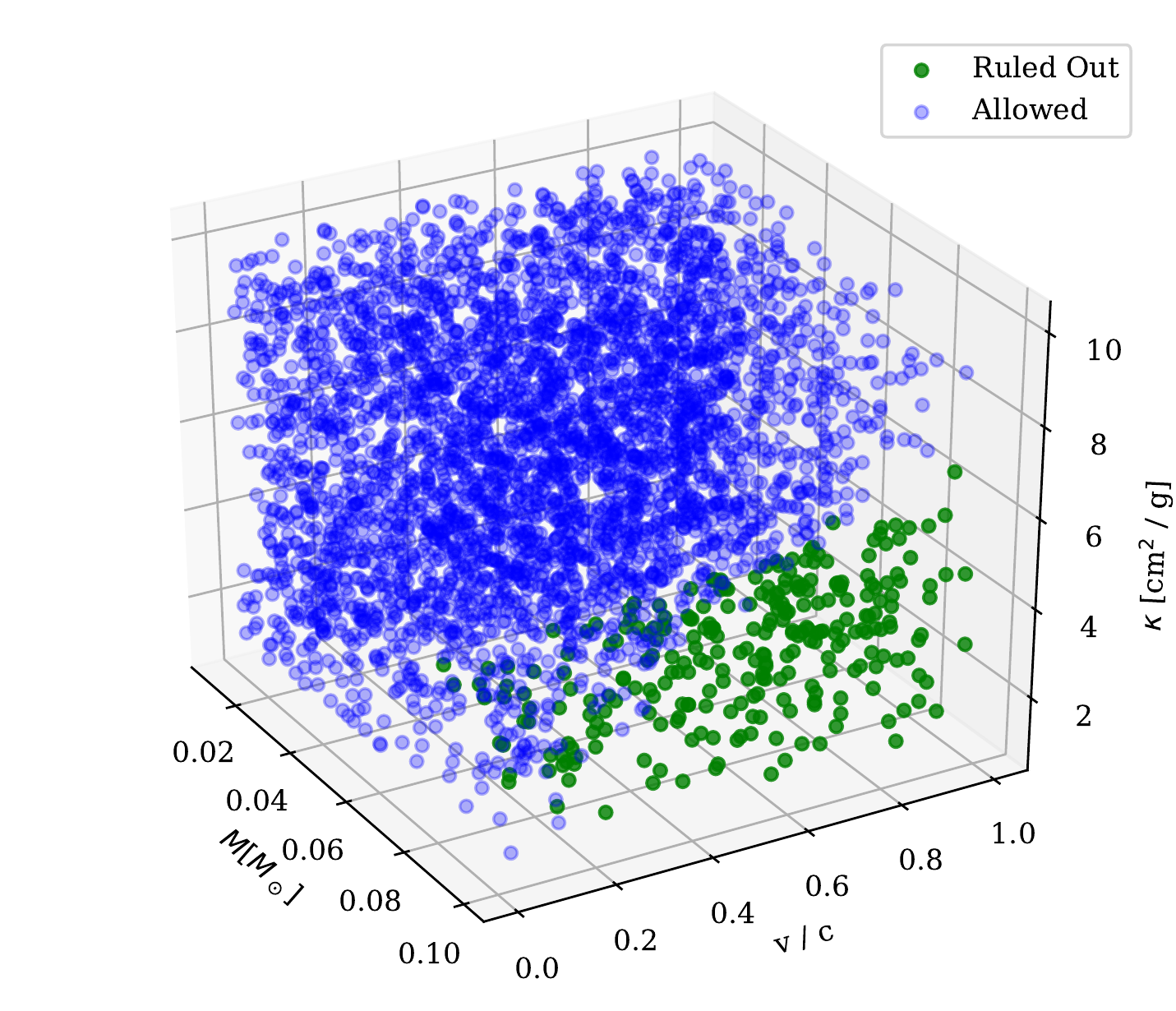}}
\caption{Parameter exploration for variations of the red kilonova model shown in Figure~\ref{fig:lightcurve}. The green points represent the models the satisfy Equation~\ref{eq2}, which we can rule out based on our $i$-band upper limits. 
\label{fig:parameters}}
\end{center}
\end{figure}

In Figure~\ref{fig:lightcurve} we show a model with parameters identical to the red component of GW170817 from the model shown in \cite{Villar+17b} ($M_{\rm ej} = 0.011$ M$_\odot$, $v_{\rm ej} = 0.137c$, $\kappa = 10$ cm$^2$ g$^{-1}$). We further explore the parameter space of allowed and ruled out models for the red/purple kilonova models, since a blue component would not be expected in a NS--BH merger. We sample 5000 random models in which we allow $M_{\rm ej}$ to vary from 0.01 to 0.1 M$_\odot$, $v_{\rm ej}$ from 0.0 to $1.0c$, and $\kappa$ from 1 to 10 cm$^2$ g$^{-1}$. In Figure~\ref{fig:parameters} we show the set of models that are ruled out by our $i$-band upper limits. We can rule out models that are brighter than either of our $3\sigma$ upper limits shown in Figure~\ref{fig:lightcurve}. The ruled out models satisfy the following equation:

\begin{equation} \label{eq2}
-0.124 \frac{\kappa}{cm^2 g^{-1}} + 11.3 \frac{M_{\rm ej}}{M_\odot} + 0.886 \frac{v_{\rm ej}}{c} > 1
\end{equation}

Finally, it is possible that an NS--BH merger can launch a relativistic jet as in short GRBs \citep{Berger2014,Paschalidis+15}.  The currently available GW information does not provide any insight on the inclination angle of the binary, so here we compare to both on- and off-axis models using the median properties of short GRB afterglows \citep{Fong+15} in the {\tt BOXFIT} software package \citep{vanEerten+11}.  We find that an on-axis afterglow can be ruled out by our observations, while a model with an off-axis viewing angle of $\gtrsim 15^\circ$ will be dimmer than about 24 mag, beyond the limits of our search.

We therefore conclude that for the region covered by our search we can generally rule out an optical counterpart similar to or brighter than GW170817, as well as an on-axis jet typical of cosmological short GRBs.  However, we cannot rule out potential scenarios such as low-mass and lanthanide-rich kilonova, or an off-axis jet.  We note again that current parameter estimation indicates no matter outside of the remnant's horizon so it is possible that S190814bv did not produce any EM radiation. 

\section{Discussion and Conclusions}
\label{sec:conc}

We presented \textit{Magellan} follow-up observations of the NS--BH merger candidate S190814bv. Our search targeted 96 galaxies in the 50\% probability region, and did not yield potential counterparts to a median limiting magnitude of $i\approx 22.2$.  We estimate that our search covered about 75\% of the integrated galaxy luminosity within the targeted region, leading to about 35\% of the effective probability for S190814bv.  Within this effective volume we can rule out the presence of an optical counterpart similar to or brighter than GW170817, as well as an on-axis afterglow typical of cosmological short GRBs.  We cannot rule out dimmer but relevant models such as a lanthanide-rich kilonova with an ejecta mass of \mbox{$0.01$ M$_\odot$}, or an off-axis jet.

We stress that to robustly rule out the presence of an optical counterpart to S190814bv for this range of models requires coverage of the full localization volume to a limiting magnitude of $\approx 25$ mag.  This is well beyond the reach of any search reported via the GCN circulars.  Thus, the existence of optical emission from S190814bv is likely to remain ambiguous.

From the point of view of GW information for S190814bv, it is unclear whether this is an NS--BH or BH--BH merger since the LVC definition of a neutron star as an object with $<3$ M$_\odot$ actually allows for both possibilities, depending on the actual mass cutoff for a neutron star. For the purpose of optimizing future EM follow-up of potential NS--BH mergers we urge the LVC to also provide the probability that the lighter binary member has a mass of $<$2 M$_\odot$, a much better indicator of a true neutron star nature than the current $<$3 M$_\odot$ definition.

In either case, the LVC has indicated that the probability for matter outside the final BH remnant is negligible ($<1\%$) suggesting that this is either an NS--BH merger with a high mass ratio and/or negligible black hole spin (in which case the neutron star was not disrupted outside the black hole horizon) or a BH--BH merger.  If the latter is the case, then our limit on optical emission (over 35\% of the probability volume) is about 1.8 times deeper than the limit on optical emission from the BH--BH merger GW170814 (at $z\approx 0.12$; \citealt{Doctor+17}).

\acknowledgements

The Berger Time-Domain Group is supported in part by NSF grant AST-1714498 and NASA grant NNX15AE50G. P.S.C.\ is grateful for support provided by NASA through the NASA Hubble Fellowship grant \#HST-HF2-51404.001-A awarded by the Space Telescope Science Institute, which is operated by the Association of Universities for Research in Astronomy, Inc., for NASA, under contract NAS 5-26555. V.A.V.\ acknowledges support from a Ford Foundation Dissertation Fellowship. W.F. and K.P. acknowledge support by the National Science Foundation under grant Nos. AST-1814782 and AST-1909358. This paper includes data gathered with the 6.5 m \textit{Magellan} Telescopes located at Las Campanas Observatory, Chile. Boxfit code was supported in part by NASA through grant NNX10AF62G issued through the Astrophysics Theory Program and by the NSF through grant AST-1009863. This research has made use of NASA’s Astrophysics Data System.

\facilities{ADS, \textit{Magellan} (IMACS)}
\software{Astropy \citep{astropy}, \texttt{BOXFIT} \citep{vanEerten+11}, \textit{healpy} \citep{healpy}, HOTPANTS \citep{Becker2015}, \texttt{ligo.skymap}, \citep{ligo.skymap, Singer2016}, Matplotlib \citep{matplotlib}, extinction \citep{Barbary16}, NumPy \citep{numpy}, PyGCN \citep{singer_python_2019}}

\bibliography{references}

\begin{longrotatetable}
\begin{deluxetable*}{lccccCCCC}
\tablecaption{Log of \textit{Magellan} Follow-up Observations \label{tab:magellan}}
\tablecolumns{9}
\tablehead{\colhead{Name} & \colhead{R.A.} & \colhead{Decl.} & \colhead{Date} & \colhead{UT} & \colhead{MJD} & \colhead{Redshift} & \colhead{$M_B$} & \colhead{Limiting Mag.\tablenotemark{a}}}
\startdata
2MASXJ00494172$-$2503029 & \ra{00}{49}{41}{70} & \dec{-25}{03}{02}{9} & 2019 Aug 16 & 08:15:37.9 & 58711.34418 & 0.0581 & -19.77 & 22.14 \\ 
PGC7877                & \ra{00}{51}{29}{90} & \dec{-24}{38}{33}{0} & 2019 Aug 16 & 08:21:39.0 & 58711.34837 & 0.0612 & -20.62 & 22.15 \\ 
PGC777629              & \ra{00}{51}{17}{80} & \dec{-25}{33}{13}{6} & 2019 Aug 16 & 08:26:24.7 & 58711.35167 & 0.0606 & -18.59 & 22.25 \\ 
PGC3235511             & \ra{00}{49}{30}{60} & \dec{-25}{16}{19}{6} & 2019 Aug 16 & 08:29:33.4 & 58711.35385 & 0.0585 & -18.68 & 22.42 \\ 
2MASXJ00485495$-$2504100 & \ra{00}{48}{55}{00} & \dec{-25}{04}{10}{1} & 2019 Aug 16 & 08:32:12.9 & 58711.35569 & 0.0540 & -20.04 & 21.80 \\ 
PGC3235463             & \ra{00}{53}{17}{70} & \dec{-25}{11}{08}{2} & 2019 Aug 16 & 08:35:01.0 & 58711.35765 & 0.0577 & -18.55 & 21.78 \\ 
PGC786964              & \ra{00}{49}{53}{50} & \dec{-24}{42}{25}{9} & 2019 Aug 16 & 08:37:56.0 & 58711.35968 & 0.0524 & -20.06 & 22.06 \\ 
ESO474-035             & \ra{00}{52}{41}{60} & \dec{-25}{44}{01}{9} & 2019 Aug 16 & 08:40:45.9 & 58711.36163 & 0.0605 & -20.96 & 22.23 \\ 
PGC786999              & \ra{00}{53}{04}{60} & \dec{-24}{42}{15}{8} & 2019 Aug 16 & 08:43:55.6 & 58711.36383 & 0.0519 & -19.80 & 21.48 \\ 
PGC787067              & \ra{00}{52}{15}{40} & \dec{-24}{41}{55}{4} & 2019 Aug 16 & 08:47:00.8 & 58711.36597 & 0.0505 & -19.86 & 22.24 \\ 
PGC3235913             & \ra{00}{51}{36}{60} & \dec{-25}{56}{31}{9} & 2019 Aug 16 & 08:49:49.6 & 58711.36793 & 0.0582 & -19.38 & 20.72 \\ 
PGC3235862             & \ra{00}{53}{24}{90} & \dec{-25}{49}{36}{5} & 2019 Aug 16 & 08:52:35.4 & 58711.36985 & 0.0579 & -19.55 & 21.77 \\ 
PGC777373              & \ra{00}{50}{52}{40} & \dec{-25}{34}{37}{4} & 2019 Aug 16 & 08:55:27.0 & 58711.37184 & 0.0507 & -19.49 & 22.42 \\ 
PGC773232              & \ra{00}{51}{10}{50} & \dec{-25}{57}{15}{0} & 2019 Aug 16 & 08:58:20.0 & 58711.37384 & 0.0623 & -20.71 & 21.89 \\ 
PGC2864                & \ra{00}{49}{01}{50} & \dec{-23}{48}{40}{7} & 2019 Aug 16 & 09:01:26.7 & 58711.37600 & 0.0525 & -21.18 & 22.16 \\ 
PGC3235517             & \ra{00}{48}{24}{80} & \dec{-25}{35}{44}{5} & 2019 Aug 16 & 09:04:14.4 & 58711.37794 & 0.0607 & -20.30 & 22.13 \\ 
PGC3235518             & \ra{00}{48}{22}{50} & \dec{-25}{36}{05}{0} & 2019 Aug 16 & 09:07:01.9 & 58711.37987 & 0.0550 & -18.97 & 22.30 \\ 
PGC198197              & \ra{00}{48}{21}{90} & \dec{-25}{07}{36}{5} & 2019 Aug 16 & 09:09:53.7 & 58711.38186 & 0.0661 & -21.43 & 22.17 \\ 
PGC773004              & \ra{00}{50}{12}{40} & \dec{-25}{58}{30}{6} & 2019 Aug 16 & 09:12:49.4 & 58711.38390 & 0.0608 & -19.96 & 21.39 \\ 
PGC198196              & \ra{00}{47}{28}{90} & \dec{-25}{26}{26}{4} & 2019 Aug 16 & 09:15:42.1 & 58711.38590 & 0.0594 & -21.36 & 22.54 \\ 
PGC797191              & \ra{00}{48}{42}{80} & \dec{-23}{46}{23}{1} & 2019 Aug 16 & 09:18:30.3 & 58711.38785 & 0.0568 & -20.47 & 21.96 \\ 
2MASXJ00530427$-$2610148 & \ra{00}{53}{04}{30} & \dec{-26}{10}{14}{9} & 2019 Aug 16 & 09:21:19.0 & 58711.38980 & 0.0545 & -20.11 & 22.20 \\ 
ESO474-026             & \ra{00}{47}{07}{50} & \dec{-24}{22}{14}{3} & 2019 Aug 16 & 09:24:18.2 & 58711.39188 & 0.0263 & -22.02 & 21.54 \\ 
IC1587                 & \ra{00}{48}{43}{30} & \dec{-23}{33}{42}{1} & 2019 Aug 16 & 09:27:10.7 & 58711.39387 & 0.0442 & -21.86 & 22.13 \\ 
PGC2998                & \ra{00}{51}{18}{80} & \dec{-26}{10}{05}{0} & 2019 Aug 16 & 09:30:10.4 & 58711.39595 & 0.0635 & -20.98 & 22.44 \\ 
PGC773198              & \ra{00}{50}{32}{90} & \dec{-25}{57}{25}{9} & 2019 Aug 16 & 09:33:01.0 & 58711.39793 & 0.0653 & -19.69 & 22.13 \\ 
PGC3235917             & \ra{00}{51}{34}{10} & \dec{-26}{04}{25}{7} & 2019 Aug 16 & 09:35:51.5 & 58711.39990 & 0.0653 & -18.82 & 19.98 \\ 
PGC3235955             & \ra{00}{49}{51}{00} & \dec{-25}{51}{42}{8} & 2019 Aug 16 & 09:38:39.9 & 58711.40184 & 0.0653 & -18.57 & 19.39 \\ 
PGC769203              & \ra{00}{51}{12}{20} & \dec{-26}{18}{47}{0} & 2019 Aug 16 & 09:41:26.9 & 58711.40377 & 0.0587 & -20.38 & 22.19 \\ 
PGC133702              & \ra{00}{48}{58}{30} & \dec{-25}{41}{36}{4} & 2019 Aug 16 & 09:44:16.8 & 58711.40574 & 0.0658 & -20.67 & 21.80 \\ 
2MASXJ00511861$-$2620430 & \ra{00}{51}{18}{60} & \dec{-26}{20}{43}{0} & 2019 Aug 16 & 09:47:05.2 & 58711.40770 & 0.0551 & -19.95 & 22.04 \\ 
PGC3235434             & \ra{00}{55}{09}{20} & \dec{-25}{27}{20}{5} & 2019 Aug 16 & 09:51:01.0 & 58711.41043 & 0.0516 & -20.63 & 22.27 \\ 
PGC774472              & \ra{00}{47}{52}{60} & \dec{-25}{50}{29}{0} & 2019 Aug 16 & 09:53:51.3 & 58711.41240 & 0.0550 & -20.11 & 22.16 \\ 
PGC783013              & \ra{00}{54}{37}{70} & \dec{-25}{04}{01}{6} & 2019 Aug 16 & 09:56:41.9 & 58711.41436 & 0.0499 & -20.46 & 22.26 \\ 
PGC798968              & \ra{00}{50}{34}{50} & \dec{-23}{37}{06}{8} & 2019 Aug 16 & 10:00:06.7 & 58711.41674 & 0.0512 & -20.51 & 21.65 \\ 
PGC773149              & \ra{00}{51}{15}{80} & \dec{-25}{57}{39}{2} & 2019 Aug 16 & 10:02:58.3 & 58711.41873 & 0.0667 & -19.56 & 21.95 \\ 
PGC771842              & \ra{00}{51}{31}{40} & \dec{-26}{04}{38}{0} & 2019 Aug 16 & 10:05:47.4 & 58711.42068 & 0.0656 & -20.08 & 21.53 \\ 
PGC3235965             & \ra{00}{49}{33}{00} & \dec{-25}{53}{50}{3} & 2019 Aug 16 & 10:08:42.4 & 58711.42271 & 0.0661 & -18.89 & 20.44 \\ 
PGC198252              & \ra{00}{47}{07}{60} & \dec{-25}{39}{38}{6} & 2019 Aug 16 & 10:11:35.0 & 58711.42471 & 0.0598 & -20.16 & 21.61 \\ 
PGC2694                & \ra{00}{46}{10}{40} & \dec{-24}{39}{00}{7} & 2019 Aug 16 & 10:14:28.9 & 58711.42671 & 0.0495 & -20.91 & 21.99 \\ 
PGC772937              & \ra{00}{51}{03}{50} & \dec{-25}{58}{56}{6} & 2019 Aug 16 & 10:17:03.2 & 58711.42851 & 0.0676 & -19.21 & 22.12 \\ 
ESO474-041             & \ra{00}{54}{24}{30} & \dec{-25}{27}{50}{6} & 2019 Aug 16 & 10:19:37.4 & 58711.43029 & 0.0506 & -21.42 & 22.22 \\ 
PGC772456              & \ra{00}{54}{21}{20} & \dec{-26}{01}{24}{3} & 2019 Aug 16 & 10:22:13.5 & 58711.43209 & 0.0496 & -19.15 & 22.24 \\ 
PGC198243              & \ra{00}{45}{35}{00} & \dec{-24}{14}{54}{7} & 2019 Aug 16 & 10:24:51.4 & 58711.43392 & 0.0516 & -19.89 & 21.75 \\ 
PGC2875                & \ra{00}{49}{14}{70} & \dec{-23}{51}{30}{8} & 2019 Aug 16 & 10:30:01.7 & 58711.43751 & 0.0440 & -20.68 & 21.91 \\ 
PGC3235867             & \ra{00}{52}{59}{00} & \dec{-26}{03}{03}{5} & 2019 Aug 17 & 08:08:38.9 & 58712.33933 & 0.0670 & -19.46 & 22.08 \\ 
PGC2875                & \ra{00}{49}{14}{70} & \dec{-23}{51}{30}{8} & 2019 Aug 17 & 08:13:23.8 & 58712.34263 & 0.0440 & -20.68 & 21.96 \\ 
IC1588                 & \ra{00}{50}{57}{70} & \dec{-23}{33}{28}{8} & 2019 Aug 17 & 08:16:19.3 & 58712.34466 & 0.0540 & -20.67 & 21.81 \\ 
PGC792107              & \ra{00}{47}{05}{30} & \dec{-24}{14}{19}{3} & 2019 Aug 17 & 08:18:56.4 & 58712.34648 & 0.0650 & -20.92 & 22.10 \\ 
PGC133715              & \ra{00}{49}{46}{00} & \dec{-26}{26}{34}{9} & 2019 Aug 17 & 08:21:31.2 & 58712.34828 & 0.0543 & -21.21 & 21.81 \\ 
PGC3123                & \ra{00}{53}{11}{80} & \dec{-26}{05}{38}{3} & 2019 Aug 17 & 08:24:10.3 & 58712.35012 & 0.0456 & -20.58 & 22.19 \\ 
PGC773323              & \ra{00}{49}{52}{20} & \dec{-25}{56}{46}{5} & 2019 Aug 17 & 08:26:42.2 & 58712.35188 & 0.0679 & -20.14 & 21.78 \\ 
PGC768565              & \ra{00}{51}{20}{50} & \dec{-26}{22}{16}{0} & 2019 Aug 17 & 08:29:18.0 & 58712.35368 & 0.0501 & -19.00 & 22.42 \\ 
PGC771948              & \ra{00}{52}{41}{90} & \dec{-26}{04}{04}{3} & 2019 Aug 17 & 08:31:50.0 & 58712.35544 & 0.0682 & -19.40 & 21.54 \\ 
PGC78883               & \ra{00}{53}{57}{70} & \dec{-24}{32}{35}{3} & 2019 Aug 17 & 08:35:44.3 & 58712.35815 & 0.0658 & -20.81 & 21.92 \\ 
PGC142558              & \ra{00}{49}{19}{70} & \dec{-26}{28}{35}{0} & 2019 Aug 17 & 08:38:17.0 & 58712.35992 & 0.0567 & -20.44 & 20.08 \\ 
PGC3235498             & \ra{00}{51}{17}{20} & \dec{-25}{32}{01}{3} & 2019 Aug 17 & 08:40:49.6 & 58712.36168 & 0.0728 & -19.49 & 21.51 \\ 
PGC766121              & \ra{00}{53}{26}{20} & \dec{-26}{35}{59}{3} & 2019 Aug 17 & 08:43:22.7 & 58712.36345 & 0.0624 & -20.51 & 21.80 \\ 
PGC769032              & \ra{00}{52}{22}{30} & \dec{-26}{19}{44}{9} & 2019 Aug 17 & 08:45:57.6 & 58712.36524 & 0.0679 & -19.26 & 21.81 \\ 
PGC3231                & \ra{00}{54}{49}{10} & \dec{-26}{22}{16}{5} & 2019 Aug 17 & 08:48:28.4 & 58712.36699 & 0.0531 & -21.34 & 22.20 \\ 
PGC787272              & \ra{00}{45}{01}{90} & \dec{-24}{40}{54}{2} & 2019 Aug 17 & 08:51:02.6 & 58712.36877 & 0.0514 & -19.50 & 21.94 \\ 
PGC765201              & \ra{00}{50}{40}{80} & \dec{-26}{40}{36}{8} & 2019 Aug 17 & 08:53:41.4 & 58712.37061 & 0.0633 & -18.75 & 22.25 \\ 
PGC3235993             & \ra{00}{48}{45}{90} & \dec{-25}{57}{58}{2} & 2019 Aug 17 & 08:56:18.6 & 58712.37243 & 0.0700 & -19.20 & 21.47 \\ 
PGC198205              & \ra{00}{50}{09}{70} & \dec{-23}{16}{48}{2} & 2019 Aug 17 & 08:59:23.5 & 58712.37457 & 0.0575 & -20.92 & 22.00 \\ 
PGC2798981             & \ra{00}{55}{12}{90} & \dec{-26}{19}{51}{5} & 2019 Aug 17 & 09:02:22.6 & 58712.37664 & 0.0533 & -19.99 & 20.76 \\ 
PGC76845               & \ra{00}{54}{53}{50} & \dec{-26}{22}{52}{3} & 2019 Aug 17 & 09:04:53.2 & 58712.37839 & 0.0579 & -19.67 & 22.11 \\ 
PGC133703              & \ra{00}{45}{53}{30} & \dec{-23}{46}{20}{9} & 2019 Aug 17 & 09:07:27.0 & 58712.38017 & 0.0526 & -21.06 & 21.77 \\ 
PGC133716              & \ra{00}{49}{27}{60} & \dec{-26}{32}{17}{9} & 2019 Aug 17 & 09:10:05.3 & 58712.38200 & 0.0504 & -21.39 & 20.74 \\ 
PGC198201              & \ra{00}{50}{33}{10} & \dec{-23}{17}{43}{8} & 2019 Aug 17 & 09:12:46.7 & 58712.38387 & 0.0532 & -20.77 & 22.26 \\ 
PGC3235988             & \ra{00}{49}{03}{00} & \dec{-26}{37}{07}{1} & 2019 Aug 17 & 09:15:44.3 & 58712.38593 & 0.0585 & -18.81 & 21.45 \\ 
2MASXJ00455322$-$2346498 & \ra{00}{45}{53}{20} & \dec{-23}{46}{49}{9} & 2019 Aug 17 & 09:18:16.2 & 58712.38769 & 0.0591 & -19.64 & 22.13 \\ 
PGC1337                & \ra{00}{53}{54}{50} & \dec{-24}{04}{37}{3} & 2019 Aug 17 & 09:20:51.3 & 58712.38948 & 0.0471 & -20.76 & 22.22 \\ 
PGC801954              & \ra{00}{50}{44}{70} & \dec{-23}{20}{19}{8} & 2019 Aug 17 & 09:23:21.8 & 58712.39122 & 0.0593 & -19.81 & 21.51 \\ 
PGC2993                & \ra{00}{51}{14}{00} & \dec{-26}{27}{40}{0} & 2019 Aug 17 & 09:25:56.0 & 58712.39301 & 0.0449 & -20.52 & 22.19 \\ 
PGC3236054             & \ra{00}{46}{17}{20} & \dec{-25}{49}{26}{4} & 2019 Aug 17 & 09:28:35.9 & 58712.39485 & 0.0598 & -19.98 & 22.13 \\ 
PGC198217              & \ra{00}{49}{23}{10} & \dec{-26}{30}{27}{0} & 2019 Aug 17 & 09:31:11.0 & 58712.39666 & 0.0665 & -20.94 & 21.28 \\ 
PGC805757              & \ra{00}{46}{53}{00} & \dec{-23}{02}{00}{2} & 2019 Aug 17 & 09:33:46.2 & 58712.39845 & 0.0559 & -20.20 & 21.65 \\ 
PGC198242              & \ra{00}{45}{03}{60} & \dec{-25}{01}{11}{2} & 2019 Aug 17 & 09:36:32.4 & 58712.40037 & 0.0614 & -20.62 & 22.08 \\ 
2MASXJ00502560$-$2434315\tablenotemark{b} & \ra{00}{50}{25}{61} & \dec{-24}{34}{31}{5} & 2019 Aug 17 & 09:38:54.0 & 58712.40201 & 0.0392 & -21.91 & 22.06 \\ 
PGC3264                & \ra{00}{55}{13}{50} & \dec{-26}{19}{16}{5} & 2019 Aug 17 & 09:39:06.3 & 58712.40215 & 0.0417 & -20.54 & 21.79 \\ 
PGC3083                & \ra{00}{52}{35}{90} & \dec{-26}{45}{03}{4} & 2019 Aug 17 & 09:41:40.8 & 58712.40394 & 0.0469 & -18.74 & 21.92 \\ 
PGC3235508             & \ra{00}{50}{23}{70} & \dec{-25}{38}{58}{9} & 2019 Aug 17 & 09:44:24.8 & 58712.40583 & 0.0738 & -20.14 & 21.49 \\ 
PGC79603               & \ra{00}{54}{43}{40} & \dec{-23}{52}{38}{7} & 2019 Aug 17 & 09:46:59.3 & 58712.40763 & 0.0564 & -20.93 & 21.28 \\ 
PGC198221              & \ra{00}{49}{32}{50} & \dec{-26}{32}{18}{9} & 2019 Aug 17 & 09:49:39.1 & 58712.40948 & 0.0670 & -19.85 & 21.89 \\ 
PGC774512              & \ra{00}{47}{42}{60} & \dec{-25}{50}{15}{3} & 2019 Aug 17 & 09:52:11.7 & 58712.41124 & 0.0706 & -19.30 & 22.46 \\ 
PGC3235964             & \ra{00}{49}{34}{20} & \dec{-26}{17}{16}{1} & 2019 Aug 17 & 09:55:03.3 & 58712.41323 & 0.0709 & -19.17 & 21.27 \\ 
PGC783349              & \ra{00}{56}{06}{70} & \dec{-25}{02}{08}{2} & 2019 Aug 17 & 09:58:20.0 & 58712.41551 & 0.0633 & -19.75 & 21.86 \\ 
PGC796342              & \ra{00}{45}{04}{00} & \dec{-23}{51}{02}{9} & 2019 Aug 17 & 10:00:56.4 & 58712.41731 & 0.0442 & -19.96 & 21.92 \\ 
PGC768185              & \ra{00}{50}{38}{90} & \dec{-26}{24}{23}{2} & 2019 Aug 17 & 10:03:29.2 & 58712.41909 & 0.0707 & -21.20 & 21.85 \\ 
PGC798818              & \ra{00}{50}{54}{40} & \dec{-23}{37}{54}{8} & 2019 Aug 17 & 10:06:19.0 & 58712.42105 & 0.0701 & -21.25 & 22.50 \\ 
PGC278                 & \ra{00}{47}{28}{20} & \dec{-23}{01}{22}{8} & 2019 Aug 17 & 10:08:54.2 & 58712.42285 & 0.0457 & -21.20 & 22.38 \\ 
PGC101138              & \ra{00}{55}{13}{10} & \dec{-24}{02}{38}{5} & 2019 Aug 17 & 10:11:31.6 & 58712.42466 & 0.0458 & -19.85 & 22.21 \\ 
PGC77801               & \ra{00}{45}{49}{60} & \dec{-25}{31}{17}{7} & 2019 Aug 17 & 10:14:31.6 & 58712.42675 & 0.0650 & -18.53 & 22.13 \\ 
PGC3235531             & \ra{00}{45}{52}{00} & \dec{-25}{29}{04}{6} & 2019 Aug 17 & 10:17:03.6 & 58712.42851 & 0.0664 & -19.23 & 22.12 \\ 
PGC781464              & \ra{00}{46}{10}{40} & \dec{-25}{12}{47}{7} & 2019 Aug 17 & 10:19:36.1 & 58712.43028 & 0.0703 & -19.51 & 21.55 \\ 
PGC769778              & \ra{00}{49}{22}{50} & \dec{-26}{15}{32}{9} & 2019 Aug 17 & 10:22:07.0 & 58712.43203 & 0.0719 & -19.08 & 22.08 \\ 
PGC769446              & \ra{00}{56}{08}{80} & \dec{-26}{17}{28}{2} & 2019 Axug 17 & 10:27:14.3 & 58712.43558 & 0.0582 & -19.08 & 22.16 \\ 
\enddata
\tablenotetext{a}{\footnotesize All of the reported magnitudes are corrected for Milky Way extinction.}
\tablenotetext{b}{\footnotesize GCN circular 25382 accidentally omitted this galaxy.}
\end{deluxetable*}
\end{longrotatetable}

\end{document}